\documentclass[11pt]{article} 
\usepackage{amssymb,amsmath} 
\begin{document} 
\begin{titlepage} 
\title{Solutions of the Gaudin Equation and Gaudin Algebras} 
\author{A.~B. Balantekin\footnote{baha@physics.wisc.edu}\\ 
\small{  Department of Physics, University of Wisconsin}\\ 
              \small{ Madison, Wisconsin 53706 USA } \\ \\ 
T.~Dereli\footnote{ tdereli@ku.edu.tr}\\ 
\small{Department of Physics, Ko\c{c} University}\\ 
\small{34450 Sar{\i}yer, \.{I}stanbul, Turkey} \\  \\ 
Y. Pehlivan \footnote{yamac@physics.wisc.edu}\\ 
\small{Department of Mathematics , \.{I}zmir University of Economics}\\ 
\small{35330 \.{I}zmir, Turkey}} 
\date{11 March 2005} 
\maketitle 
\begin{abstract} 
\noindent Three well-known solutions of the Gaudin equation are 
obtained under a set of standard assumptions. By relaxing one of 
these assumptions we introduce a class of mutually commuting 
Hamiltonians based on a different solution of the Gaudin equation. 
Application of the algebraic Bethe ansatz technique to diagonalize 
these Hamiltonians reveals a new infinite dimensional complex Lie 
algebra. 
\end{abstract} 
 
\vskip 1cm 
 
 
\end{titlepage} 
 
\section{Introduction} 
 
\noindent The Bardeen-Cooper-Schrieffer (BCS) pairing model 
Hamiltonian was diagonalized by Richardson  in 1963 in an 
algebraic way \cite{Richardson}. The step operators introduced in 
Richardson's solution belong to an infinite dimensional complex 
Lie algebra which is usually referred to as the rational Gaudin 
algebra. It is one of the three algebras which emerged from 
Gaudin's work during the 1970's \cite{gaudin1} who introduced a 
new class of integrable models starting with the following 
operators: 
\begin{equation}\label{hGaudin} 
h_i=\sum_{{{j=1}\atop{j\neq i}}}^N \sum_{\alpha=0}^2 w_{ij}^\alpha 
t_i^\alpha t_j^\alpha 
\end{equation} 
where $w_{ij}^\alpha$ are  complex numbers to be determined and 
$t_i^\alpha$ are the generators of $N$, mutually commuting $SU(2)$ 
algebras. The latter obey the standard $SU(2)$ commutation 
relations: 
\begin{equation} 
\label{su2} [t_i^+,t_j^-]=2\delta_{ij} t_j^0,\quad [t_i^0,t_j^\pm 
]=\pm \delta_{ij}t_j^\pm \quad , \quad i,j = 1,2,...,N 
\end{equation} 
where $t_i^\pm \equiv t_i^1 \pm it_i^2$. Gaudin showed that the 
operators in (\ref{hGaudin}) mutually commute 
\begin{equation}\label{hcommute} 
[h_i,h_j]=0 
\end{equation} 
if and only if the coefficients satisfy the equations: 
\begin{equation}\label{Gaudin's Equation} 
w_{ij}^\alpha w_{jk}^\gamma + w_{ji}^\beta w_{ik}^\gamma - 
w_{ik}^\alpha w_{jk}^\beta = 0 
\end{equation} 
for all distinct triples $(i,j,k)$ and for all permutations of the 
upper indices $(0,1,2)$. There are three solutions to equation 
(\ref{Gaudin's Equation}) under the following assumptions: i) The 
coefficients $w_{ij}^\alpha$ are antisymmetric under the exchange 
of the indices $i$ and $j$ 
\begin{equation}\label{antisymmetry} 
w_{ij}^\alpha+w_{ji}^\alpha=0 , 
\end{equation} 
ii) each coefficient $w_{ij}^\alpha$ can be expressed as a 
function of the difference between two real parameters $u_i$ and 
$u_j$, and iii) each $h_i$ commutes with the operator $T=\sum_j 
t_j^0$  so that the $z$ component of the total $SU(2)$ is 
conserved. Gaudin's solutions are given as follows: 
\begin{equation}\label{rationalsolution} 
w_{ij}^\alpha = \frac{1}{u_i-u_j} \quad\mbox{for}\quad 
\alpha=0,1,2, 
\end{equation} 
\begin{equation}\label{trigonometricsolution} 
w_{ij}^0 = p\cot [p(u_i-u_j)] \quad , \quad w_{ij}^{1} = 
w_{ij}^{2} = \frac{p}{\sin[p(u_i-u_j)]} , 
\end{equation} 
\begin{equation}\label{hyperbolicsolution} 
w_{ij}^0 = p\coth[p(u_i-u_j)] \quad , \quad w_{ij}^{1} = 
w_{ij}^{2} = \frac{p}{\sinh[p(u_i-u_j)]} , 
\end{equation} 
which are commonly referred to as rational, trigonometric and 
hyperbolic solutions, respectively. Here, $p$ is a real parameter 
\footnote{These solutions are not completely unrelated. Gaudin 
equation (\ref{Gaudin's Equation}) is satisfied for all complex 
values of $p$. But only for the real and pure imaginary values of 
$p$ Hamiltonians in (\ref{hGaudin}) are Hermitian. On the other 
hand, substituting $ip$ in place of $p$, one can convert 
trigonometric and hyperbolic solutions into each other. For this 
reason, we restrict ourselves with the real values of $p$. Also 
note that in the limit $p \to 0$ trigonometric and hyperbolic 
solutions go to the rational solution.}. 
 
The operator which is obtained by substituting the rational 
solution into (\ref{hGaudin}) is called rational Gaudin magnet 
Hamiltonian: 
\begin{equation}\label{hrational} 
h^{(r)}_i=\sum_{{{j=1}\atop{j\neq i}}}^N 
\frac{\overrightarrow{t}_i \cdot \overrightarrow{t}_j}{u_i-u_j} 
\end{equation} 
where 
\begin{equation*} 
\overrightarrow{t}_i \cdot \overrightarrow{t}_j=t^0_it^0_j+t^1_i 
t^1_j + t^2_i t^2_j =t^0_it^0_j+\frac{1}{2}(t^+_i t^-_j + t^-_i 
t^+_j). 
\end{equation*} 
Similarly, substituting the trigonometric and hyperbolic solutions 
into (\ref{hGaudin}) we obtain 
\begin{equation}\label{htrigonometric} 
h^{(t)}_i = \sum_{{{j=1}\atop{j\neq i}}}^N \left ( 
p\cot{[p(u_i-u_j)]}t_i^0t_j^0 + 
\frac{p}{2}\frac{t_i^+t_j^-+t_i^-t_j^+}{\sin[p(u_i-u_j)]} \right ) 
\end{equation} 
\begin{equation}\label{hhyperbolic} 
h^{(h)}_i=\sum_{{{j=1}\atop{j\neq i}}}^N \left ( 
p\coth{[p(u_i-u_j)]}t_i^0t_j^0 + 
\frac{p}{2}\frac{t_i^+t_j^-+t_i^-t_j^+}{\sinh[p(u_i-u_j)]} \right 
) 
\end{equation} 
which are called trigonometric and hyperbolic Gaudin magnet 
Hamiltonians, respectively. 
 
Application of the algebraic Bethe ansatz method to the Gaudin 
magnets yields the rational, trigonometric and hyperbolic Gaudin 
algebras \cite{gaudin2,ush}.  These are infinite dimensional 
complex Lie algebras which are related to particular solutions of 
the classical Yang-Baxter equation \cite{faddeev}. As a result, 
each algebra admits a one-parameter family of mutually commuting 
Hamiltonians $H(\lambda)$. Here $\lambda$ is a complex parameter 
which is usually referred to as the spectral parameter. These 
operators may be identified as the integrals of motion of a 
quantum system as well as the traces of transfer matrices of a 
vertex model. Main objective in both cases is to diagonalize them 
simultaneously and this is achieved by either functional or 
algebraic Bethe ansatz techniques \cite{Faddeev:1996iy}. 
Richardson-Gaudin methods have recently found many applications in 
quantum many-body physics \cite{Dukelsky:2004re}. 
 
 
\section{Further Solutions of the Gaudin Equation} 
 
In this paper we investigate a different solution of the Gaudin 
equation  (\ref{Gaudin's Equation}). Under the three assumptions 
listed in the introduction, all possible solutions are enumerated 
above. To find new solutions we need to relax one or more of these 
assumptions. We keep the constraints (ii) and (iii) but generalize 
the constraint (i) to 
\begin{equation}\label{qsymmetry} 
w_{ij}^\alpha+w_{ji}^\alpha=-2q , 
\end{equation} 
where $q$ is a real parameter\footnote{In general, (\ref{Gaudin's 
Equation}) is satisfied for all complex values of $q$. But only 
for real values of $q$ one obtains Hermitian Hamiltonians in 
(\ref{hGaudin}).}. 
 This solution is given by 
\begin{equation}\label{newsolution} 
w_{ij}^\alpha = q\coth[q(u_i-u_j)]-q \quad \mbox{for}\quad 
\alpha=0,1,2. 
\end{equation} 
 The operators which are 
obtained by substituting this new solution in (\ref{hGaudin}) will 
be denoted by $h^{(q)}_i$: 
\begin{equation}\label{hnew} 
h^{(q)}_i=\sum_{{{j=1}\atop{j\neq i}}}^N \left ( 
q\coth[q(u_i-u_j)]-q \right )\overrightarrow{t}_i \cdot 
\overrightarrow{t}_j . 
\end{equation} 
Since (\ref{newsolution}) is a solution of Gaudin equation 
(\ref{Gaudin's Equation}), these operators mutually commute: 
\begin{equation}\label{hqcommute} 
[h^{(q)}_i,h^{(q)}_j]=0 , 
\end{equation} 
In the limit  $q\to 0$, the new solution given by 
(\ref{newsolution}) approaches  Gaudin's rational solution 
(\ref{rationalsolution}). As a result, 
\begin{equation} 
\lim_{q \to 0} h_i^{(q)}=h_i^{(r)}. 
\end{equation} 
 
The solution presented here is a different solution of Gaudin 
equation and the Hamiltonians given by (\ref{hnew}) for $q \neq 0$ 
cannot be obtained from the rational, trigonometric or hyperbolic 
Gaudin magnet Hamiltonians. To see that, let us assume for a 
moment that there exists an operator $S$ such that 
\begin{equation}\label{assumption} 
Sh_i^{(*)}S^{-1}=h_i^{(q)} 
\end{equation} 
where $h_i^{(*)}$ represents any one of the Hamiltonians given in 
equations (\ref{hrational})-(\ref{hhyperbolic}). Suppose we take 
the sum of both sides of (\ref{assumption}): 
\begin{equation}\label{sum} 
\sum_{i=1}^N Sh_i^{(*)}S^{-1}=\sum_{i=1}^N h_i^{(q)}. 
\end{equation} 
Antisymmetry of (\ref{antisymmetry}) implies 
\begin{equation} 
\sum_{i=1}^N h_i^{(*)}=0 
\end{equation} 
for rational, trigonometric and hyperbolic magnet Hamiltonians. 
Hence the sum on the left hand side is equal to zero. 
 On the other hand, the sum on the right hand side of 
(\ref{sum}) is not equal to zero. Instead the new condition given 
in (\ref{qsymmetry}) implies 
\begin{equation} 
\sum_{i=1}^N h_i^{(q)}=-q\sum_{{i,j=1}\atop{i \neq j}}^N 
\overrightarrow{t}_i \cdot \overrightarrow{t}_j  , 
\end{equation} 
i.e. an operator $S$ satisfying (\ref{assumption}) does not exist 
for $q \neq 0$. 
 
 
\section{Gaudin Algebras} 
 
If one applies the algebraic Bethe ansatz technique to diagonalize 
the Gaudin magnet Hamiltonians which correspond to the solutions 
(\ref{rationalsolution}), ({\ref{trigonometricsolution}), or 
(\ref{hyperbolicsolution}), one uses the step operators belonging 
to the rational, trigonometric or hyperbolic Gaudin algebras, 
respectively. Below we will show that if the algebraic Bethe 
ansatz method is employed to diagonalize the  Hamiltonian of 
(\ref{hnew}), then a different algebra emerges in a completely 
analogous way. Before introducing this algebra, however, we would 
like to outline the basic features of rational Gaudin algebra and 
its relation to the rational Gaudin magnet Hamiltonian, as an 
illustration of the technique. Trigonometric and hyperbolic Gaudin 
algebras and their relation to the trigonometric and hyperbolic 
Gaudin magnet Hamiltonians follow very similar lines. 
 
Rational Gaudin algebra is generated by three families of 
operators $J^+(\lambda),J^-(\lambda)$ and $J^0(\lambda)$ 
parameterized by a complex number $\lambda$. Commutators are given 
by 
\begin{equation*} 
[J^+(\lambda),J^-(\mu)]=2\frac{J^0(\lambda)-J^0(\mu)}{\lambda-\mu}, 
\end{equation*} 
\begin{equation}\label{GAUDIN_ALGEBRA} 
[J^0(\lambda),J^{\pm}(\mu)]=\pm\frac{J^{\pm}(\lambda)-J^{\pm}(\mu)} 
                                                  {\lambda-\mu}, 
\end{equation} 
\begin{equation*} 
[J^0(\lambda),J^0(\mu)]=[J^{\pm}(\lambda),J^{\pm}(\mu)]=0. 
\end{equation*} 
Commutators of $J^+(\lambda), J^-(\lambda)$ and $J^0(\lambda)$ at 
the same value of the complex parameter are given by taking the 
limit $\mu \rightarrow \lambda$. From these commutation relations 
it is easy to show that 
\begin{equation}\label{DEFINE_H} 
H(\lambda)=J^0(\lambda)J^0(\lambda)+\frac{1}{2}J^+(\lambda)J^-(\lambda)+ 
           \frac{1}{2}J^-(\lambda)J^+(\lambda) 
\end{equation} 
form a one-parameter family of mutually commuting operators: 
\begin{equation}\label{H_COMMUTATORS} 
[H(\lambda),H(\mu)]=0. 
\end{equation} 
Starting from a lowest weight vector and using $J^+(\lambda)$ as 
step operators, one can diagonalize these operators 
simultaneously. Lowest weight vector $|0>$ by definition satisfies 
\begin{equation}\label{LOWEST_WEIGTH_STATE} 
J^-(\lambda)|0>=0,\quad\mbox{and}\quad 
J^0(\lambda)|0>=W(\lambda)|0> 
\end{equation} 
for every $\lambda$. Here $W(\lambda)$ is a complex-valued 
function. The state $|0>$ itself is an eigenvector of 
$H(\lambda)$: 
\begin{equation} 
H(\lambda)|0>=E_0(\lambda)|0> 
\end{equation} 
with the eigenvalue 
\begin{equation} 
E_0(\lambda)=W(\lambda)^2-W'(\lambda) 
\end{equation} 
where the prime  denotes derivative with respect to $\lambda$. In 
addition 
\begin{equation} 
|\xi_1,\xi_2,\dots,\xi_n>\equiv J^+(\xi_1) J^+(\xi_2)\dots 
J^+(\xi_n)|0> 
\end{equation} 
is an eigenvector of $H(\lambda)$ if the quantities 
$\xi_1,\xi_2,\dots, \xi_n$ satisfy the set of Bethe ansatz 
equations 
\begin{equation}\label{Bethe Ansatz Equation} 
W(\xi_\alpha)=\sum_{ {\beta=1}\atop{(\beta\neq\alpha)} }^n 
\frac{1}{\xi_\alpha-\xi_\beta} \quad \mbox{for} \quad 
\alpha=1,2,\dots,n. 
\end{equation} 
(For an analytic solution of these equations see e.g. References 
\cite{Balantekin:2004yf} and \cite{Balantekin:2004yf2}). The 
corresponding eigenvalues will be 
\begin{equation} 
E_n(\lambda)=E_0(\lambda)-2\sum_{\alpha=1}^n 
\frac{W(\lambda)-W(\xi_\alpha)}{\lambda-\xi_\alpha}. 
\end{equation} 
 
There exists a realization of the rational Gaudin algebra in terms 
of the SU(2) generators of  (\ref{su2}), given by 
\begin{equation} \label{REALIZATION} 
J^{0}(\lambda) = 
\sum_{i=1}^N\frac{t_i^0}{u_i-\lambda}\quad\mbox{and}\quad 
J^{\pm}(\lambda)=\sum_{i=1}^N \frac{t_i^\pm}{u_i-\lambda}. 
\end{equation} 
Here $u_1,u_2,\dots,u_N$ are arbitrary real numbers which are all 
different from each other and $N$ is a nonnegative 
integer\footnote{In general $u_1,u_2,\dots,u_N$ can be complex and 
they  need not be different from each other. But in most physical 
applications we are interested in those realizations for which 
$u_1,u_2,\dots,u_N$ are real and different from each other. 
Reality guarantees that $J^+(\lambda)^\dagger=J^-(\lambda^*)$ and 
$J^0(\lambda)^\dagger=J^0(\lambda^*)$}. Assuming that the 
eigenvalues of the Casimir operator of the $j^{th}$ $SU(2)$ are 
$s_j(s_j +1)$, the corresponding $W(\lambda)$ is given by 
\begin{equation} 
W(\lambda)=\sum_{i=1}^N \frac{-s_i}{u_i-\lambda}. 
\end{equation} 
In this realization $H(\lambda)$ is given by 
\begin{equation}\label{H_IN_REALIZATION} 
H(\lambda)= 
\sum_{i,j=1}^N\frac{\overrightarrow{t_i}\cdot\overrightarrow{t_j}} 
                        {(u_i-\lambda)(u_j-\lambda)}. 
\end{equation} 
We see that $H(\lambda)$ has simple poles on the real axis. 
Residues of $H(\lambda)$ at the points $\lambda=u_i$ are 
proportional to the rational Gaudin magnet Hamiltonians given in 
(\ref{hrational}): 
\begin{equation} 
\label{33} 
-\frac{1}{2}\mbox{Res}\left\{H(\lambda))\right\}_{\lambda=u_i}=h^{(r)}_i 
\end{equation} 
Equation (\ref{H_COMMUTATORS}) implies 
\begin{equation}\label{h_COMMUTATORS} 
[H(\lambda),h^{(r)}_i]=0, \quad\quad\mbox{and}\quad\quad 
[h^{(r)}_i,h^{(r)}_j]=0. 
\end{equation} 
As a result, rational Gaudin magnet Hamiltonians $h^{(r)}_i$ are 
automatically diagonalized together with $H(\lambda)$. It is easy 
to read off the eigenvalues of the rational Gaudin magnet 
Hamiltonians from the eigenvalues of $H(\lambda)$ as follows: 
\begin{equation} 
h_i^{(r)} |0>=E^{(r)}_{i,0}|0>, 
\end{equation} 
where 
\begin{equation} 
E^{(r)}_{i,0}=\sum_{{j=1}\atop{(j\neq i)}}^N 
            \frac{s_is_j}{u_i-u_j} 
\end{equation} 
and 
\begin{equation} 
h_i^{(r)} 
|\xi_1,\xi_2,\dots,\xi_n>=E^{(r)}_{i,n}|\xi_1,\xi_2,\dots,\xi_n> 
\end{equation} 
where 
\begin{equation} 
E^{(r)}_{i,n}=E^{(r)}_{i,0}-s_i\sum_{\alpha=1}^n 
          \frac{1}{u_i-\xi_\alpha}. 
\end{equation} 
Here $|0>$ is the lowest weight state and $\xi_\alpha$ are the 
solutions of the Bethe ansatz equation (\ref{Bethe Ansatz 
Equation}).

We next search for an operator $H^{(q)}(\lambda)$, the residue of 
which gives us the operators $h_i^{(q)}$ (cf. Equation 
(\ref{33})). We wish to write such an operator using an algebra 
similar to that in (\ref{GAUDIN_ALGEBRA}). Below we show that such 
an algebra exists and its generators $J_q^{0}(\lambda)$, 
$J_q^{+}(\lambda)$ and $J_q^{-}(\lambda)$ satisfy the commutation 
relations given as 
\begin{equation*} 
[J_q^+(\lambda) , J_q^-(\mu)]= 
2q\frac{J_q^0(\lambda)-J_q^0(\mu)}{\tanh{[q(\lambda-\mu)]}}+ 
2q\left(J_q^0(\lambda)+ J_q^0(\mu)\right), 
\end{equation*} 
\begin{equation}\label{New Algebra} 
[J_q^0(\lambda),J_q^{\pm}(\mu)] = 
{\pm}q\frac{J_q^{\pm}(\lambda)-J_q^{\pm}(\mu)}{\tanh{[q(\lambda-\mu)]}} 
\pm q \left( J_q^{\pm}(\lambda)+J_q^{\pm}(\mu) \right), 
\end{equation} 
\begin{equation*} 
[J_q^0(\lambda),J_q^0(\mu)]=[J_q^{\pm}(\lambda),J_q^{\pm}(\mu)]=0. 
\end{equation*} 
We observe that, in the limit $q\to 0$, commutators in  (\ref{New 
Algebra}) approach the commutators of Gaudin algebra given by 
(\ref{GAUDIN_ALGEBRA}). It should nevertheless be emphasized that 
the algebra of  (\ref{New Algebra}) is not a q-deformed algebra, 
but an ordinary Lie algebra. We can show that 
\begin{equation}\label{DEFINE_H_q} 
H^{(q)}(\lambda)=J_q^0(\lambda)J_q^0(\lambda)+ 
\frac{1}{2}J_q^+(\lambda)J_q^-(\lambda)+\frac{1}{2} 
J_q^-(\lambda)J_q^+(\lambda) 
\end{equation} 
form a one-parameter commutative family: 
\begin{equation}\label{H_q_COMMUTATOR} 
[H^{(q)}(\lambda),H^{(q)}(\mu)]=0 . 
\end{equation} 
We can simultaneously diagonalize them starting from a lowest 
weight vector satisfying 
\begin{equation}\label{LOWEST_WEIGTH_STATE_q} 
J_q^-(\lambda)|0>=0, \quad J_q^0(\lambda)|0> = W_q(\lambda)|0> 
\end{equation} 
where $W_q(\lambda)$ is a complex valued function. It can be shown 
that $|0>$  is an eigenvector of $H^{(q)}(\lambda)$ with the 
eigenvalue 
\begin{equation}\label{LOWEST_WEIGTH_EIGENVALUE_q} 
E^{(q)}_{0}(\lambda)=W_q(\lambda)^2-W_q'(\lambda)-2q W_q(\lambda). 
\end{equation} 
In general, one can write a Bethe ansatz form for the eigenvectors 
of $H^{(q)}(\lambda)$ as 
\begin{equation}\label{BETHE_ANSATZ_q} 
|\xi_1,\xi_2,\dots,\xi_n>\equiv J_q^+(\xi_1)J_q^+(\xi_2) \dots 
J_q^+(\xi_n)|0>. 
\end{equation} 
For this to be an eigenvector of $H^{(q)}(\lambda)$, the complex 
numbers $\xi_1,\xi_2,\dots,\xi_n$ must satisfy the following 
system of equations: 
\begin{equation}\label{BETHE_ANSATZ_EQUATION_q} 
W_q(\xi_\alpha)=\sum_{{\beta=1}\atop{(\beta \neq \alpha)}}^n 
q\left(\coth{[q(\xi_\alpha-\xi_\beta)]}-1\right)\quad\quad 
\mbox{for} \quad \alpha=1,2,\dots,n. 
\end{equation} 
In this case, eigenvalue associated with the state 
$|\xi_1,\xi_2,\dots,\xi_n>$ is given by 
\begin{equation}\label{BETHE_ANSATZ_EIGENVALUE_q} 
E^{(q)}_{n}(\lambda)=E^{(q)}_{0}(\lambda) - 2 \sum_{\alpha=1}^n 
q\left(\coth{[q(\lambda-\xi_\alpha)]}-1\right)\left( 
W_q(\lambda)-W_q(\xi_\alpha) \right). 
\end{equation} 
This algebra admits two realizations in terms of the $SU(2)$ 
generators of  (\ref{su2}). These are given by 
\begin{equation} \label{REALIZATION_q} 
J_q^{\pm,0}(\lambda)=\sum_{i=1}^N 
q\left(\coth{[q(u_i-\lambda)]}+1\right) t_i^{\pm,0} 
\end{equation} 
and 
\begin{equation} 
J_q^{\pm,0}(\lambda)=\sum_{i=1}^N 
q\left(\tanh{[q(u_i-\lambda)]}+1\right) t_i^{\pm,0}. 
\end{equation} 
In the limit $q\to 0$, the first realization goes to the 
realization of the rational Gaudin algebra given by 
(\ref{REALIZATION}) whereas the second realization vanishes. 
 
In the first realization, Equation (\ref{REALIZATION_q}), the 
operator $H^{(q)}(\lambda)$ becomes 
\begin{equation} 
H^{(q)}(\lambda) = \sum_{i,j=1}^N 
\left(q\coth{[q(u_i-\lambda)]}+q\right)\left(q\coth{[q(u_j-\lambda)]}+q\right) 
\overrightarrow{t_i}\cdot\overrightarrow{t_j}. 
\end{equation} 
We see that $H^{(q)}(\lambda)$ has simple poles on real axis at 
the points $\lambda=u_i$. It is easy to show that $-1/2$ times the 
residue of $H^{(q)}(\lambda)$ at the point $\lambda=u_i$ is 
\begin{equation}\label{2t} 
-\frac{1}{2}\mbox{Res}\left\{H_q(\lambda))\right\}_{\lambda=u_i}= 
h_{i}^{(q)}-q\overrightarrow{t}_i\cdot\overrightarrow{t}_i. 
\end{equation} 
Since $\overrightarrow{t_i}\cdot\overrightarrow{t_i}$ commutes 
with every $h_{j}^{(q)}$ and $H^{(q)}(\lambda)$, 
(\ref{H_q_COMMUTATOR}) implies 
\begin{equation}\label{hq_COMMUTATORS} 
[H^{(q)}(\lambda),h^{(q)}_i]=0, \quad\quad\mbox{and}\quad\quad 
[h^{(q)}_i,h^{(q)}_j]=0. 
\end{equation} 
As a result, the operators $h^{(q)}_i$ are automatically 
diagonalized together with $H^{(q)}(\lambda)$. As in the case of 
the rational Gaudin algebra, one can read off the eigenvalues of 
$h^{(q)}_i$ from the eigenvalues of $H^{(q)}(\lambda)$: 
\begin{equation} 
h_i^{(q)} |0>=E^{(q)}_{0,i}|0>, 
\end{equation} 
where 
\begin{equation} 
E^{(q)}_{0,i}=\sum_{{j=1}\atop{(j\neq i)}}^N 
s_is_jq(\coth{[q(u_i-u_j)]}-1) 
\end{equation} 
and 
\begin{equation} 
h_i^{(q)} 
|\xi_1,\xi_2,\dots,\xi_n>=E^{(q)}_{n,i}|\xi_1,\xi_2,\dots,\xi_n> 
\end{equation} 
where 
\begin{equation} 
E^{(q)}_{n,i}=E^{(q)}_{0,i}-s_i\sum_{\alpha=1}^n 
q(\coth{[q(u_i-\xi_\alpha)]}-1). 
\end{equation} 
Details of the derivation of these eigenvalues are given in the 
Appendix. 
\section{Linear r-Matrix Structure} 
 
The rational, trigonometric and hyperbolic Gaudin algebras have a 
linear $r$-matrix structure. This means that one can write down 
the commutators of these algebras in the following matrix form: 
\begin{equation}\label{L EQUATION} 
[L(\lambda)\otimes I,I\otimes L(\mu)] + 
[r(\lambda-\mu),L(\lambda)\otimes I+I\otimes L(\mu)]=0 
\end{equation} 
where $I$ is the $2\times 2$ identity matrix, $r(\lambda-\mu)$ is 
the r-matrix described below and $L(\lambda)$ is given by 
\begin{equation}\label{L matrix} 
L(\lambda)=\begin{pmatrix} 
  J^0(\lambda) & J^+(\lambda) \\ 
  J^+(\lambda) & -J^0(\lambda) 
\end{pmatrix}. 
\end{equation} 
$J^0(\lambda),J^+(\lambda)$ and $J^-(\lambda)$ are generators of 
the rational, trigonometric or hyperbolic Gaudin algebras.  In 
(\ref{L EQUATION}), the term $[L(\lambda)\otimes I,I\otimes 
L(\mu)]$ is equal to a $4 \times 4$ matrix whose elements are 
various commutators between the generators 
$J^0(\lambda),J^+(\lambda)$ and $J^-(\lambda)$. On the other hand 
$r(\lambda-\mu)$ is a $Mat_2(\mathbb{C}) \otimes 
Mat_2(\mathbb{C})$ matrix valued function of $\lambda-\mu$ which 
carries the information about the structure constants of the 
algebra. For instance, for the rational Gaudin algebra, the 
$r$-matrix is given by 
\begin{equation}\label{Rational r matrix} 
r(\lambda-\mu)=\frac{1}{\lambda-\mu}P 
\end{equation} 
where $P$ is the permutation matrix on 
$\mathbb{C}^2\otimes\mathbb{C}^2$, which is equal to 
\begin{equation}\label{Permutation Matrix} 
P=\begin{pmatrix}1&0&0&0\\ 0&0&1&0\\ 
0&1&0&0\\0&0&0&1\end{pmatrix}. 
\end{equation} 
It is customary to introduce the following three mappings 
\begin{eqnarray}\label{maps} 
&\varphi^{12}:a\otimes b\longrightarrow a\otimes b\otimes I,\\ 
&\varphi^{13}:a\otimes b\longrightarrow a\otimes I\otimes b,\\ 
&\varphi^{23}:a\otimes b\longrightarrow I\otimes a\otimes b, 
\end{eqnarray} 
where $a,b$ are $2 \times 2$ matrices and $I$ is the $2 \times 2$ 
identity matrix. Then $r^{ij}$ are defined as follows: 
\begin{equation}\label{mapsto} 
r^{12}=\varphi^{12}[r(\lambda-\mu)],\quad\quad 
r^{13}=\varphi^{13}[r(\lambda-\sigma)],\quad\quad 
r^{23}=\varphi^{23}[r(\mu-\sigma)]. 
\end{equation} 
Since $r$-matrix carries information about the structure constants 
of the algebra, Jacobi identity leads to the equality 
\begin{equation}\label{CYBE} 
[r^{13},r^{23}]+[r^{12},r^{13}]+[r^{12},r^{23}] =0 
\end{equation} 
which is the well known classical Yang-Baxter equation (for 
solutions of this equation see \cite{Belavin&Drinfel'd}, for a 
review see \cite{Hoppe}, \cite{Jurco}). Equation (\ref{CYBE}) 
guarantees the mutual commutativity of the operators $H(\lambda)$ 
which can now be written as 
\begin{equation}\label{Trace of L} 
H(\lambda)=\frac{1}{2}\mbox{Tr}\left\{L(\lambda)^2\right\}. 
\end{equation} 
Trigonometric and hyperbolic Gaudin algebras also have the 
$r$-matrix structure described above and their $r$-matrices are 
also solutions of the classical Yang-Baxter equation. 
 
In order to see if one can repeat the same procedure for the 
algebra we introduced in (39), we study a matrix of the form 
\begin{equation}\label{L q matrix} 
L_q(\lambda)=\begin{pmatrix} 
  J_q^0(\lambda) & J_q^+(\lambda) \\ 
  J_q^+(\lambda) & -J_q^0(\lambda) 
\end{pmatrix}. 
\end{equation} 
We then substitute it into the equation 
\begin{equation}\label{Lq  trial Equation} 
[L_q(\lambda)\otimes I,I\otimes L_q(\mu)] + 
[r(\lambda-\mu),L_q(\lambda)\otimes I+I\otimes L_q(\mu)]=0 , 
\end{equation} 
assuming the existence of such an $r$-matrix. One can compute the 
left-hand side of this equation and then set all the components 
equal to zero to find the components $r^{ij}$ of this $r$-matrix. 
For instance, the $(21)$ and $(13)$ elements are equal to 
\begin{eqnarray*} 
&\left( r_{22}-r_{11}+q\coth \left[ q\left( \lambda -\mu \right) 
\right] -q\right) J_{q}^{-}\left( \mu \right)\\ 
&+\left( r_{23}-q\coth \left[ q\left( \lambda -\mu \right) \right] 
-q\right) J_{q}^{-}\left( \lambda \right) -r_{41}J_{q}^{+}\left( 
\lambda \right) +2r_{21}J_{q}^{0}\left( \mu \right) 
\end{eqnarray*} 
and 
\begin{eqnarray*} 
&\left( r_{11}-r_{33}-q\coth \left[ q\left( \lambda -\mu \right) 
\right] -q\right) J_{q}^{+}\left( \lambda \right)\\ 
&-\left( r_{23}-q\coth \left[ q\left( \lambda -\mu \right) \right] 
+q\right) J_{q}^{+}\left( \mu \right) +r_{14}J_{q}^{-}\left( \mu 
\right) -2r_{13}J_{q}^{0}\left( \lambda \right), 
\end{eqnarray*} 
respectively. Setting these elements equal to zero requires 
setting the coefficients of all the algebra elements equal to 
zero. The coefficient of $J_q^-(\lambda)$ in the first equation 
and the coefficient of $J_q^+(\mu)$ in the second equation gives 
\begin{equation} 
r_{23}-q\coth[q(\lambda-\mu)]-q=0 \quad\mbox{and}\quad 
r_{23}-q\coth[q(\lambda-\mu)]+q=0, 
\end{equation} 
respectively. But these equations are incompatible unless $q=0$. 
Repeating the same calculations for the other components, one can 
conclude that there is no $r$-matrix which satisfies (\ref{Lq 
trial Equation}). Instead of the $L_q(\lambda)$ matrix in (\ref{L 
q matrix}), one may try to find a more general form of 
$L_q(\lambda)$ matrix for which an $r$-matrix can be found to 
satisfy  (\ref{Lq trial Equation}). But we were unable to find 
such an $L_q(\lambda)$ matrix \footnote{ For instance, for a 
matrix of the form 
\begin{equation*} 
L_q(\lambda)= 
\begin{pmatrix} 
\alpha J^{0}(\lambda) & \beta J^{+}(\lambda)+\gamma J^{-}\lambda)\\ 
\beta J^{-}(\lambda)-\gamma J^{+}(\lambda) & -\alpha 
J^{0}(\lambda) 
\end{pmatrix}, 
\end{equation*} 
$Tr[L_q(\lambda)L_q^\dagger(\lambda)]$ is proportional to 
$H^{(q)}(\lambda)$ when $\alpha^2=\beta^2+\gamma^2$. However, one 
can similarly show that there is no solution to (\ref{Lq trial 
Equation}) for this $L_q(\lambda)$ either.}. On the other hand we 
were able to show that one can write the commutators of the 
algebra introduced in (\ref{New Algebra}) in the following form: 
\begin{equation}\label{Lq Equation} 
[L_q(\lambda)\otimes I, I\otimes L_q(\mu)]+[r_q(\lambda-\mu), 
L_q(\lambda)\otimes I]+[r_{-q}(\lambda-\mu),I\otimes L_q(\mu)]=0. 
\end{equation} 
Here $L_q(\lambda)$ is given by (\ref{L q matrix}) and the 
$r_q$-matrix is given by 
\begin{equation}\label{rq matrix} 
r_q(\lambda-\mu)=(q\coth{[q(\lambda-\mu)]}+q)P. 
\end{equation} 
$P$ is the permutation matrix introduced in (\ref{Permutation 
Matrix}). As in the case of the $r$-matrices of rational, 
trigonometric and hyperbolic Gaudin algebras, the $r_q$-matrix 
carries information about the structure constants of the algebra 
of  (\ref{New Algebra}). $r_q^{ij}$ matrices, defined by similar 
maps as in Equations (\ref{maps})-(\ref{mapsto}),  satisfy  the 
following equation: 
\begin{equation}\label{qCYBE} 
[r_{-q}^{13},r_q^{23}]+[r_{-q}^{12},r_q^{23}]+[r_{-q}^{12},r_q^{13}] 
= 0 . 
\end{equation} 
We see that the $r$-matrix structure of the algebra introduced in 
 (\ref{New Algebra}) is not the same as the $r$-matrix structures 
of the rational, trigonometric and hyperbolic Gaudin algebras. 
Nevertheless, it shares with the other three algebras, the crucial 
property of admitting a one-parameter family of mutually commuting 
operators $H^{(q)}(\lambda)$. Similar to (\ref{Trace of L}), one 
can write this operator as the trace of the square of 
$L_q(\lambda)$: 
\begin{equation} 
H^{(q)}(\lambda)=\frac{1}{2}\mbox{Tr} 
\left\{L_q(\lambda)^2\right\}. 
\end{equation} 
Note that in the limit $q \to 0$, the algebra of  (\ref{New 
Algebra}) approaches  the rational Gaudin algebra. In this limit 
both $r_q$ and $r_{-q}$ go to the $r$-matrix of the rational 
Gaudin algebra given in  (\ref{Rational r matrix}) and from 
(\ref{qCYBE})  we recover  the classical Yang-Baxter equation 
(\ref{CYBE}). 
 
\section{Conclusion} 
 
In this article, by relaxing one of the conditions imposed by 
Gaudin, we presented a different solution to the  Gaudin equation 
and wrote down the corresponding set of mutually commuting 
Hamiltonians. We also identified the related infinite dimensional 
Lie algebra. This algebra allows a one parameter family of 
mutually commuting Hamiltonians $H^{(q)}(\lambda)$ parameterized 
by a complex spectral parameter. We diagonalized these 
Hamiltonians using the algebraic Bethe ansatz method and 
constructed the  eigenvectors and eigenvalues. The procedure is 
parallel to the other known Gaudin algebras.

A different generalization of the Gaudin algebras was recently 
introduced in Ref.\cite{Ortiz:2004yv} that is based on the 
rational, trigonometric, and hyperbolic solutions of the Gaudin 
equations. Our generalization is based on a different solution of 
the Gaudin equation obtained under different assumptions. 
 
 
\section*{Acknowledgement} 
 
\noindent This  work   was supported in  part  by   the  U.S. 
National Science Foundation Grants No. INT-0352192 and PHY-0244384 
at the University of  Wisconsin, and  in  part by  the  University 
of Wisconsin Research Committee   with  funds  granted by the 
Wisconsin Alumni  Research Foundation. We also acknowledge support 
through NSF-TUBITAK Joint Research Project TBAG-U/84(103T113). 
 
 
\section*{Appendix: Computation of the Eigenvalues} 
 
 To see that the lowest weight state given by (\ref{LOWEST_WEIGTH_STATE_q}) 
is an eigenstate of $H^{(q)}(\lambda)$ we first use the 
commutators (\ref{New Algebra}) to write $H^{(q)}(\lambda)$ in the 
following form: 
\begin{eqnarray} 
H^{(q)}(\lambda)&=&J^0(\lambda)J^0(\lambda)+ 
J^+(\lambda)J^-(\lambda)+\frac{1}{2}[J^-(\lambda),J^+(\lambda)]\nonumber\\ 
&=&J^0(\lambda)J^0(\lambda)+J^+(\lambda)J^-(\lambda)\\ 
&&-\frac{1}{2}\lim_{\mu\to\lambda}2q\left( 
\frac{J^0(\lambda)-J^0(\mu)}{\tanh{[q(\lambda-\mu)]}} 
+J^0(\lambda)+J^0(\mu)\right).\nonumber 
\end{eqnarray} 
Then the action of $H^{(q)}(\lambda)$ on the lowest weight state 
is 
\begin{equation} 
H^{(q)}(\lambda)|0> 
=\left(W(\lambda)^2-W'(\lambda)-2qW(\lambda)\right)|0>. 
\end{equation} 
We see that the lowest weight state is an eigenstate of 
$H^{(q)}(\lambda)$ with the eigenvalue $E_0(\lambda)$ given by 
(\ref{LOWEST_WEIGTH_EIGENVALUE_q}). 
 
 Now consider the state $|\xi>=J^+(\xi)|0>$. 
Action of $H^{(q)}(\lambda)$ on $J^+(\xi)|0>$ is 
\begin{eqnarray} 
&H^{(q)}(\lambda)J^+(\xi)|0>= 
2q\left(\coth{[q(\lambda-\xi)]}-1\right)W(\xi)J^+(\lambda)|0>\\ 
&+\left(E_0(\lambda)-2qW(\lambda)\coth{[q(\lambda-\xi)]} 
+2qW(\lambda)\right)J^+(\xi)|0>.\nonumber 
\end{eqnarray} 
We see that $H^{(q)}(\lambda)J^+(\xi)|0>$ is a superposition of 
the states $J^+(\lambda)|0>$ and $J^+(\xi)|0>$. Therefore, for a 
generic $\xi$, the vector $J^+(\xi)|0>$ is not an eigenstate of 
$H^{(q)}(\lambda)$. But if $\xi$ is a root of the $W(\lambda)$ 
then the coefficient of $J^+(\lambda)|0>$ vanishes. In this case 
$J^+(\xi)|0>$ is an eigenstate of $H^{(q)}(\lambda)$ with the 
eigenvalue $E_1(\lambda)$ given by 
\begin{equation} 
E^{(q)}_1(\lambda)=E_0(\lambda)-2\left(q\coth{[q(\lambda-\xi)]} 
-q\right)W(\lambda). 
\end{equation} 
This is the energy one finds by substituting $n=1$ and $W(\xi)=0$ 
in Eqn. (\ref{BETHE_ANSATZ_EIGENVALUE_q}). 
 
For $n>1$, action of $H^{(q)}(\lambda)$ on the state 
$J^+(\xi_1)J^+(\xi_2)\dots J^+(\xi_n)|0>$ is given by 
\begin{eqnarray} 
&&H^{(q)}(\lambda)J^+(\xi_1)J^+(\xi_2)\dots J^+(\xi_n)|0>= 
J^+(\xi_1)J^+(\xi_2)\dots J^+(\xi_n)H^{(q)}(\lambda)|0>\nonumber\\ 
&&-\sum_{\alpha=1}^n 
2q\left(\coth{[q(\lambda-\xi_\alpha)]}-1\right) 
\left(\underset{\beta\neq \alpha}{\sum_{\beta=1}^n} q\left( 
\coth{[q(\xi_\alpha-\xi_\beta)]}-1\right)-W(\xi_\alpha)\right) 
\nonumber\\ 
&&\times J^+(\xi_1)\dots 
J^+(\xi_{\alpha-1})J^+(\lambda)J^+(\xi_{\alpha+1})\dots 
J^+(\xi_n)|0>\nonumber\\ 
&&+\sum_{\alpha=1}^n 
2q\left(\coth{[q(\lambda-\xi_\alpha)]}-1\right) 
\left(\underset{\beta\neq \alpha}{\sum_{\beta=1}^n} q\left( 
\coth{[q(\xi_\alpha-\xi_\beta)]}-1\right)-W(\lambda)\right)\nonumber\\ 
&&\times J^+(\xi_1)J^+(\xi_2)\dots J^+(\xi_n)|0>. 
\end{eqnarray} 
Here we see that $H^{(q)}(\lambda)J^+(\xi_1)J^+(\xi_2)\dots 
J^+(\xi_n)|0>$ is a superposition of the states 
\begin{equation} 
J^+(\xi_1)\dots 
J^+(\xi_{\alpha-1})J^+(\lambda)J^+(\xi_{\alpha+1})\dots 
J^+(\xi_n)|0> 
\end{equation} 
for $\alpha=1,2,\dots,n$ and the state 
\begin{equation} 
J^+(\xi_1)J^+(\xi_2)\dots J^+(\xi_n)|0>. 
\end{equation} 
Coefficients of the states $J^+(\xi_1)\dots J^+(\lambda)\dots 
J^+(\xi_n)|0>$ vanish when the conditions 
(\ref{BETHE_ANSATZ_EQUATION_q}) are satisfied. Consequently, 
$J^+(\xi_1)J^+(\xi_2)\dots J^+(\xi_n)|0>$ become an eigenstate of 
$H^{(q)}(\lambda)$ and its eigenvalue is given by 
(\ref{BETHE_ANSATZ_EIGENVALUE_q}).

\end{document}